\documentclass[twocolumn]{autart}    

\usepackage{epsfig}
\usepackage{amsfonts}
\usepackage{listings}
\usepackage[utf8]{inputenc}
\usepackage{amsmath,amssymb,amsfonts}
\usepackage{lmodern}
\usepackage{mathtools}
\usepackage{graphicx}
\usepackage{textcomp}
\usepackage{subfigure}
\usepackage{epstopdf}
\usepackage{dcolumn}
\usepackage{multicol}
\usepackage{enumerate}
\usepackage{color}
\usepackage{lscape}

\usepackage{hyperref}

\usepackage{algorithm}

\usepackage{stmaryrd}
\usepackage{epstopdf}
\usepackage{subfigure} 

\usepackage{multirow}
\usepackage{caption,tabularx,booktabs}

\usepackage{tikz}
\usetikzlibrary{positioning,arrows}

\newtheorem{theo}{Theorem}

\usepackage{graphicx}      

\begin{document}

\begin{frontmatter}

\title{Super-twisting based sliding mode control of hydraulic actuator without velocity state} 

\thanks[footnoteinfo]{\textcolor[rgb]{0.00,0.00,1.00}{Authors accepted manuscript, Oct 2023}}

\author[First]{Manuel A. Estrada}
\author[Second]{Michael Ruderman}
\author[First]{Leonid Fridman}

\address[First]{Facultad de Ingenier\'ia, Universidad Nacional Aut\'onoma de M\'exico (UNAM), 04510 Mexico City, Mexico. (mannyestrada94@gmail.com, lfridman@unam.mx)}
\address[Second]{Department of Engineering Sciences, University of Agder, 4879-Grimstad, Norway. (michael.ruderman@uia.no)}

\begin{abstract}
This paper provides a novel surface design and experimental
evaluation of a super-twisting algorithm (STA) based control for
hydraulic cylinder actuators. The proposed integral sliding
surface allows to track a sufficiently smooth reference without
using the velocity state which is hardly accessible in the noisy
hydraulic systems. A design methodology based on LMI's is given,
and the STA gains are designed to be adjusted by only one free
parameter. The feasibility and effectiveness of the proposed
control method are shown on a standard hydraulic test bench with
one linear degree of freedom and passive load, where a typical
motion profile is tracked with a bounded average error {below $1\%$ from the total drive effective output.}
\end{abstract}

\begin{keyword}
sliding mode control, robust control design, hydraulic actuator,
motion control
\end{keyword}

\end{frontmatter}
\section{Introduction}

Hydraulic actuators are one of the most common options to work in
various applications where high forces in harsh environments are
required for a robust operation, cf. \cite{Merritt1967}. In that
sense, motion control of such actuators remain a challenging task
due to their complex nonlinear behavior and uncertainties, cf.
e.g. \cite{Yao2000}. Notwithstanding, most of today's hydraulic
applications still rely on a few standard linear feedback
controllers, like e.g. PID, see for instance \cite{CRO2023}.

On the other hand, there are also extensions to nonlinear-type
PID controllers with extra compensators, such as for example
feed-forward compensation of the dead-zone in valves which was
shown to be useful for improving the performance, see e.g.
\cite{Liu2000}. Apart from that, there are experimental
evaluations on how the uncertainties and process noise affect the
controlled response of such systems
\cite{Niksefat2001,Alleyne2000}. One can also notice that there
exist several remarkable challenges in the implementation of such
actuators in real world applications due to environmental
conditions \cite{Ortiz2014}. More references related to the
controlled hydraulic applications can also be found in e.g.
\cite{Mattila2017}.

Over and above that, the control problem is not only attributed to
the mechanical properties of the hydraulic actuators. It is rather
hydraulic by-effects which can cause large uncertainties in the
whole system dynamics. Then, it appears imperative that the
hydro-dynamical behavior must also be taken into account, from the
control design point of view, see e.g.
\cite{Ruderman2017,Pedersen2017}.

Per se, sliding mode control has proven to be an efficient control
solution against uncertainties and external disturbances, see e.g.
\cite{Edwards1998,Edwards2001,Choi2007} as related to the design
tools used in the present work. Therefore, it appears suitable to
control also the hydraulic systems, as it was presented in some
previous works, e.g. \cite{Bonchis2001} where a sliding mode
observer was designed to estimate the so-called equivalent control
and use it to enforce the sliding mode \cite{Utkin1992}. In
\cite{Loukianov2009}, a sliding surface combined with an
$H_{\infty}$ approach was presented to robustly track a desired
trajectory. On the other hand, an application of the integral
sliding mode based approach was shown in industrial scenario in
\cite{Komsta2013}. In that work, an input-output linearization
technique with the design of an integral sliding surface was
proposed to control the pressure of a die-cushion (hydraulic)
cylinder drive, and a comparison with standard PI and PID
controllers was provided.

Although several previously mentioned solutions obtained a
sufficiently good performance, their approach relies often on the
first-order (discontinuous) sliding mode control. The use of such
discontinuous control produces the so-called \textit{chattering}
effect,
that means undesired high frequency oscillations which can damage
actuators and the plant. Note that despite chattering appears
generally as unavoidable for all-types of real sliding mode
controllers (due to additional parasitic actuator lag and
neglected sub-dynamics in feedback loop), here it is especially a
discontinuous control action which is most challenging for
hydro-mechanical actuators and plants, like control valves and
pressurized cylinders.

For substituting a discontinuous control by a continuous one,
\cite{Levant1993} introduced for instance the super-twisting
algorithm (STA). As application of STA in hydraulic systems, an
observer-based sliding mode control of low-power hydraulic
cylinder, in the case of unknown load forces, can be found for
example in \cite{Koch2016}. Also, an STA-based observer of the
large unknown load forces in hydraulic drives was presented
experimentally in \cite{ruderman2019virtual}.

Despite the above mentioned attempts and references therein, an
optimal solution to robustly track a given trajectory while
generating a continuous control signal remains an sufficiently
open problem for hydraulic systems. In particular, the lacking
measurement of the system states constitutes a large practical
issue. While the output relative displacement (i.e. stroke) of
cylinder drives is often measured by the attached (yet noisy)
linear encoders, the relative velocity remains mostly unknown. It
can also be hardly estimated (for feedback control) from the
stroke measurement with a relatively low SNR (signal-to-noise
ratio). Similarly, the measured pressure signals, if available,
contain often a high measurement and process noise.

Against the background mentioned above, the contribution of the
present work is summarized as follows.
\begin{itemize}
\item A novel integral sliding surface is proposed, such that,
the pressure state works as a virtual control for mechanical
sub-system, achieving the robust tracking of the desired
trajectory without any measurement or estimation of the velocity
signal.
\item An STA-based control law in combination with equivalent-type control
produces the required differential pressure, as a desired virtual
control, while generating the continuous control signal.
\item An easy to follow design of the control gains is provided for application.
\item The proposed approach is evaluated experimentally on a full-scale nonlinear system,
such that the feasibility of the results is shown in presence of
the complex system dynamics and uncertainties.
\end{itemize}

The structure of the paper is as follows. In section
\ref{sec:Hydradynamics}, the nonlinear dynamics of the hydraulic
cylinder actuator is summarized. Section \ref{sec:Lin} presents a
linearization of the model and formulate the control problem. In
section \ref{sec:Sol}, the solution of this problem is proposed and the control design is left for section \ref{sec:design}.  The reader may be interested in the application of an output based control law using an estimator for the missing state, therefore, in section \ref{sec:comparison} a comparison with the approach presented in \cite{oliveira2018} is presented.
The experimental evaluations are provided in section
\ref{sec:numeval}. Finally, the conclusions are drawn in section
\ref{sec:conclusions}, and the proofs are left in the Appendix.

\textbf{Notation.} For $x\in \mathbb{R}^n$, $\Vert x\Vert $
denotes the Euclidean norm and $\mathbb{C}^n$ stands for the set
of $n$-times continuous differentiable functions. The
$\text{sign}$ function is defined on $\mathbb{R}$ by $
\text{sign}(z) = \frac{z}{\vert z\vert}$ for $z \neq 0$ and
$\text{sign}(0) \in \left[-1,1 \right]$. The following notation is
used throughout the paper, for $s\in \mathbb{R}$ and $q\in
\mathbb{R}$,  $\lceil s\rfloor^q = \left\vert s\right\vert^q
\text{sign}(s)$. For a matrix $G\in \mathbb{R}^{n\times n}$,
$\lambda_{min}(G)$ (resp. $\lambda_{max}(G)$) denotes the minimum
(resp. maximum) eigenvalue of $G$. The identity matrix of an
appropriate dimension is denoted by $\mathbb{I}$. Solutions for
systems with discontinuous right-hand side are understood in the
Filippov sense \cite{filippov1988}.

\section{Dynamics of hydraulic actuator} \label{sec:Hydradynamics}

The dynamics of the hydraulic cylinder actuator is briefly
summarized for the sake of clarity. For more details on hydraulic
circuits we refer to \cite{Merritt1967}, and on those related to
the used type of linear-stroke actuators to e.g.
\cite{Ruderman2017}.

\subsection{Servo-valve approximation}

The servo-valve used to control the flow into the cylinder has the
own dynamics corresponding to electro-magnetic actuation of a
solenoid which moves the spool inside the valve. The internal
dynamics of the controlled servo-valve can be well approximated by
the second-order
\begin{equation}
 \label{eq:valve}
\ddot{\nu} + 2\zeta_v\omega_0\dot{\nu} + \omega_0^2\nu =
\omega_0^2U(t),
\end{equation}
where the output $\nu(t)$ is the relative spool position, and the
input value $U(t)$ is the applied valve control reference. The
parameters $\omega_0$ and $\zeta_v$ are the eigen-frequency and
damping ratio, respectively, of the low-level controlled
closed-loop system. The parameter values have been experimentally
identified in the work \cite{Pasolli2018} and are inline with the
valve's manufacturer data-sheet. Additionally, the relation
between the spool position and the input flow rate is not linear
due to an overlap in the spool area and saturation effects inside
the servo-valve. These output nonlinearities lead to the whole
actuator dynamics \eqref{eq:valve} augmented by
\cite{Ruderman2017}
\begin{equation}
g(t) =  \begin{cases}
c_s \text{sign}(\nu(t))\,, \quad \quad \quad \quad \vert \nu(t)\vert \geq c_s + c_d\\
\,0\,, \quad \quad \quad \quad \quad \quad \quad \quad \,\, \vert \nu(t)\vert < c_d \\
\nu(t) - c_d\text{sign}(\nu(t)) \,, \quad \text{otherwise},
\end{cases}
\end{equation}
where $g(t)$ is the valve's orifice opening which is governing the
flow through it. The constants $c_d$ and $c_s$ are the dead-zone
size and saturation level, respectively.

\subsection{Orifice and continuity equations}

As described in \cite{Ruderman2017}, with the load related
pressure between the both chambers $P = P_A - P_B$, and under the
hypothesis of a closed hydraulic circuit, that means the absolute
value of the orifice equations for both chambers are essentially
the same, one has the single orifice equation
\begin{equation}
Q = g K_f \sqrt{\frac{1}{2}\left( P_S - \text{sign}(g)P\right)}\,.
\end{equation}
Here $Q$ is the overall (load) flow rate of the hydraulic medium,
and $P_S$ is the supply pressure. $K_f$ is the valve flow
coefficient, available from the manufacturer data of the
servo-valve. Then, one can assume that the pressure in each
chamber satisfy the following
\begin{equation}
P_A = \frac{P_S + P}{2}\,\quad P_B = \frac{P_S - P}{2}\,,
\end{equation}
provided zero pressure in the tank, which is a reasonable
assumption for various hydraulic drive systems. With this in mind,
the load pressure gradient has the form
\begin{equation}
\dot{P} = \frac{4E}{V_t}\left( Q - A\dot{q} - C_LP\right) +
\delta_P(t)\,,
\end{equation}
where $V_t$ is the total volume in the hydraulic circuits, i.e.
between the pressurized inlet and zero-pressure outlet of the
control servo-valve, including the cylinder with piping. The
average effective piston area of the cylinder is $A = 0.5
\left(A_A + A_B\right)$, being $A_A$ and $A_B$ the effective
cross-sections in the chambers $A$ and $B$ respectively. The
coefficient $C_L$ corresponds to some internal leakage (if any)
between the chambers of cylinder, and $\delta_P(t)$ constitutes
the all (summarized) uncertainties in the pressure dynamics. The
latter are related to the model simplification and side-effects,
like changes in the coefficients due to varying temperature and
external effects, manufacturing tolerances, wear, and others.

\subsection{Mechanical sub-system}

The mechanical part corresponds to one linear degree of freedom
(1DOF) mechanical system, i.e. moving cylinder piston, of the form
\begin{equation}
m \ddot{q}(t) = AP - f(\dot{q}) - F_L(t)\,,
\end{equation}
where $m$ is the total mass of the moving piston rod with payload,
$F_L(t)$ is an external perturbation interfering in the
mechanical part, and $q$ is the linear displacement of cylinder.
On the other hand, a weakly known friction term can comply with
\cite{Ruderman2017}
\begin{equation}
\label{eq:friction} f(\dot{q}) = \tanh( \vartheta \dot{q})\left[
F_c + (F_s-F_c)\text{exp}\left(-\vert \dot{q}\vert^{\iota}
\chi^{-\iota} \right) \right] + \sigma \dot{q}
\end{equation}
which contains an approximation of the Coulomb friction with the
smoothing parameter $\vartheta \gg 1$ for motion reversals. The
linear viscous friction coefficient is $\sigma > 0$, the
coefficients related to the Stribeck effect are $F_s > F_c$,
$\chi>0$, and $\iota \neq 0$. Note that while \eqref{eq:friction}
provides certain structure for the friction-related perturbations,
the numerical parameter values and thus the whole $f(\cdot)$ can
remain weakly or even fully unknown during the operation.

\section{Linearized model and problem statement}\label{sec:Lin}

It is worth emphasizing, right from the beginning, that the
servo-valve dynamics is significantly faster comparing to the
hydro-mechanical sub-system and is well specified by the
manufacturer. It can be seen from the previous work
\cite{Pasolli2018} that it is at least two orders of magnitude
faster in terms of the time constants. Therefore, it is assumed
that the input $g$ is not subject to additional dynamics, cf. with
section 2.1, while the effective parasitic input dynamics (of the
valve as actuator) is respected when designing the STA control
gains, cf. \cite{Ventura2019}. Recall the orifice equation
$$
Q = \frac{g K_f}{\sqrt{2}} \sqrt{P_S - P\text{sign}(g)}
$$
and consider $\Omega \equiv \sqrt{P_S - P\text{sign}(z)}$ as an
input gaining factor that can have variations depending on the
hydro-mechanical load. By a physical reasoning that the load
pressure $P$ can not be larger than the supply pressure $P_S$, it
is satisfying $0<\epsilon_{\Omega}\leq \Omega \leq \sqrt{P_S}$.
Now, considering the operation points $g_0$ and $P_0$ of the
orifice opening and load pressure, respectively, one can linearize
the system and obtain the coefficients $C_{qp} =\left.
\frac{\partial Q}{\partial P}\right\vert_{g=g_0}$ and $C_{q} =
\frac{1}{\sqrt{2}}K_f \Omega$. Then, with the new state variables
$$
x^T = \begin{bmatrix} x_1 & x_2 & x_3 \end{bmatrix} \equiv
\begin{bmatrix} q & \dot{q}& P \end{bmatrix}
$$
one obtains the following linearized model
\begin{equation}
\begin{aligned}
\dot{x}_1 &= x_2, \\
\dot{x}_2 &= -\frac{\sigma}{m}x_2 + \frac{A}{m}x_3 + \delta_2(t,x_2) + F_L(t),\\
\dot{x}_3 &= -\frac{4 E A}{V_t}x_2 - \frac{4EC_{qp} }{V_t}x_3 + \frac{4EC_q}{V_t}u(t) + \delta_3(t)\,.
\end{aligned}
\end{equation}
where under the assumption of the servo-valve dynamics having a unitary gain and a low phase lag, the control input is supposed to be $u(t) = \nu(t)$. Here, perturbation $\delta_3(t)$ is equivalent to the perturbation in the load
pressure $\delta_P(t)$, and $\delta_2(t,x_2)$ is a vanishing
perturbation which includes also the unknown friction term
\eqref{eq:friction}. Introduce the tracking error variables as
\begin{equation}
\begin{aligned}
e_1& = x_1 - r(t),\\
e_2 & = x_2 - \dot{r}(t),\\
\eta & =\tau  x_3,
\end{aligned}
\end{equation}
where $r(t)\in \mathbb{C}^{4}$ is the reference signal to be
tracked, and $\tau>0$ is a positive scaling constant for the
measurable $x_3$ state.  Defining the perturbation terms as
 \begin{equation}
 \begin{aligned}
 \bar{\delta}_2(t,e) & =  \delta_2(t,x_2) -
\delta_2(t,\dot{r}(t))\,, \\
 \beta(t) &= \delta_2(t,\dot{r}(t)) - \ddot{r}(t) - \frac{\sigma}{m}\dot{r}(t)+ F_L(t)\,,\\
 \bar{\delta}_3(t) &= \tau \delta_3(t) - \frac{4\tau E A}{V_t}\dot{r}(t)\,,
 \end{aligned}
 \end{equation}
the error dynamics can be written as
\begin{equation}
\label{eq:12}
\begin{aligned}
\dot{e}_1 &= e_2\,, \\
\dot{e}_2 &= -\frac{\sigma}{m} e_2+ \frac{A}{\tau m} \eta +\bar{\delta}_2(t,e) + \beta(t)\,,\\
\dot{\eta} &= -\frac{4\tau E A}{V_t} e_2 - \frac{4 E C_q}{V_t}
\eta  + \frac{4\tau EC_q}{V_t}u(t) + \bar{\delta}_3(t)\,.
\end{aligned}
\end{equation}

\textbf{Control problem:} For some positive constants
$\overline{e}$ and $T_e$, it is required to enforce the error
states to a vicinity of the origin $ (e_1,e_2)=0$, \hbox { i.e. }
$\left\Vert  \left[ e_1(t), e_2(t)\right]^T \right\Vert \leq
\overline{e}$ and that: (i) for all $t>T_e$, (ii) despite the
presence of perturbations, and (iii) ensuring a continuous control
signal.

\section{Proposed control solution}
\label{sec:Sol}

The proposed control design consists of two consecutive steps:
\begin{enumerate}
\item To control the error in the mechanical sub-dynamics
$(\dot{e}_1,\dot{e}_2)$ using the internal state $\eta$ as a
virtual control input, and to compensate for perturbations
$\bar{\delta}_2(t)$.
\item To design a sliding surface $s(e_1,\eta)$, which depends only on two
measurable states, so as to enforce $\eta $ to control the
mechanical sub-system despite all perturbations by using the
super-twisting algorithm (STA).
\end{enumerate}

\begin{assum}\label{ass:1}
For all $\bigl(t,x(t)\bigr) \in \mathbb{R}_+\times \mathbb{R}^3$,
the following condition is satisfied for some known non-negative
constants $\Psi$ and $L_3$:
\begin{equation}
\begin{gathered}
\left\vert \frac{\bar{\delta}_2(t,e)}{ x_2(t)- \dot{r}(t)  } \right\vert \leq \Psi\,,  \quad
\left\vert \frac{d}{dt}\bar{\delta}_3(t)\right\vert \leq L_3\,,
\end{gathered}
\end{equation}
\end{assum}
Note that the above condition on $\bar{\delta}_2(t,x)$ is
satisfied for structured frictional perturbation in the mechanical
systems, cf. \eqref{eq:friction}.
Then one can bring the system \eqref{eq:12} into the following
form
\begin{equation}\label{eq:errordynamics}
\begin{aligned}
\dot{e}_1 &= e_2\,, \\
\dot{e}_2 &= -\frac{\sigma}{m} e_2+  \frac{A}{\tau m} \eta + \psi(t,e)e_2  + \beta(t)\,,\\
\dot{\eta} &= -\frac{4\tau E A}{V_t} e_2 - \frac{4 E C_{qp}}{V_t}
\eta  + \frac{4\tau EC_q}{V_t}u(t) + \bar{\delta}_3(t)\,,
\end{aligned}
\end{equation}
with $\psi(t,e) = \bar{\delta}_2(t,e)/e_2$. It is worth mentioning
that because of Assumption \ref{ass:1}, it is implied that
$\psi(t,e)$ is always bounded and $\lim_{e_2 \rightarrow
0}\psi(t,e)$ exists and is well defined.

In the following, the virtual control law will be designed. First,
the nominal case (i.e. stabilization) when $\beta(t) = 0$ is
considered, that is, when the reference signal is constant and
there are no external forces. In this case, it can be proved that
$e_1$ and $e_2$ converge asymptotically to zero. However, in the
case of $\beta \neq 0$, only an uniformly ultimate boundedness can
be achieved. Both results are presented separately for the sake of
clearness.

\subsection{Virtual control for mechanical sub-system}

In this section, the design of a virtual control without requiring
the velocity is presented. Assuming that the mechanical
sub-system's input is the scaled pressure state $\eta(t)$, let us
propose a virtual control design. The following virtual control is
proposed as
\begin{equation}\label{eq:v}
\eta(t) = \int v(e_1,\eta)dt - \left(\kappa+ \frac{4\tau E
A}{V_t}\right)e_1\,,
\end{equation}
where $v(e_1,\eta) =  -\gamma_1e_1 -\gamma_2\eta$, and $\kappa,
\gamma_1,\gamma_2>0$ are the design parameters.
Let us assume that $\beta(t) = 0$, and consider the new virtual state:
\begin{equation}
e_3 = \frac{A}{\tau m} \left( \int v(e_1,\eta)dt - \kappa e_1 - \frac{4\tau E A}{V_t} e_1 \right) \,.
\end{equation}
Then, the error dynamics of the mechanical sub-system can be
represented as
\begin{equation}
\begin{aligned}
\dot{e}_1 &= e_2\,, \\
\dot{e}_2 &=- \frac{A}{\tau m}\left( \frac{4\tau E A}{V_t} +\kappa\right) e_1 - \frac{\sigma}{m} e_2 \\ &+
\frac{A}{\tau m} \int v(e_1,\eta)dt+ \psi(t,e)e_2\,.
\end{aligned}
\end{equation}
Noting that $\int
v(e_1,\eta)dt = \eta + \kappa e_1 + \frac{4\tau E A}{V_t} e_1$, then $\eta
= \frac{\tau m}{A} e_3$ such that
\begin{equation}\label{eq:esystem}
\begin{aligned}
\dot{e}_1 &= e_2\,, \\
\dot{e}_2 &=-a_{22}e_2 + e_3  + \psi(t,e)e_2\,, \\
\dot{e}_3 &=  a_{23}v\left(e_1,  \frac{\tau m}{A} e_3\right) - \frac{A}{\tau m}\left( \frac{ 4\tau E A}{V_t}+
\kappa \right) e_2\,,
\end{aligned}
\end{equation}
One can recognize that the whole error dynamics system becomes
\begin{equation}
\dot{e} =( A_n - BR)e + \psi(t,e)He,
\end{equation}
with $e^T = \begin{bmatrix} e_1 & e_2 & e_3 \end{bmatrix} $ and
\begin{equation}\label{eq:matrices}
\begin{gathered}
A_n = \begin{bmatrix}
0 & 1 & 0 \\ 0 & -\frac{\sigma}{ m} & 1 \\ 0 & 0 & 0
\end{bmatrix} \, ,\quad
B = \begin{bmatrix}
0 \\ 0 \\ 1
\end{bmatrix} \,, \quad  H = \begin{bmatrix}
0 & 0 & 0 \\ 0 & 1 & 0 \\ 0 & 0 & 0
\end{bmatrix}
\end{gathered}
\end{equation}
The vector $R$ is now to be designed, noting that the poles of the
matrix $(A_n-BR)$ can always be collocated by selecting the vector
$R =\begin{bmatrix} r_1 & r_2 & r_3 \end{bmatrix} =
\begin{bmatrix} a_{23}\gamma_1 & a_{23} \left(\kappa+\alpha\right)
& \gamma_2 \end{bmatrix}$. Then it comes from controllability of
the pair $(A_n,B)$ that one can design $\gamma_1 = \frac{\tau
m}{A} r_1 $, $\gamma_2 = r_3$ and $\kappa = \frac{\tau m}{A} r_2 -
\alpha $ and therefore, assign the dynamics of the extended
sub-system via \eqref{eq:v} without requiring the velocity
variable.
\begin{theo}
Assume that $\beta(t) = 0$ and Assumption \ref{ass:1} is
fulfilled. The error variables $e(t) \rightarrow 0$ as
$t\rightarrow \infty$ if and only if for some $M=M^T>0$ the
following LMI holds:
\begin{equation}\label{eq:LyapIneq}
MA_n+A_n^TM -R^TB^TM - MBR+\Psi H^TM + \Psi MH<0\,.
\end{equation}
\end{theo}
\textit{Proof}. The proof is given in Appendix A.
\begin{rem} 
The dynamic behavior can be assigned through LMI approach so that
to satisfy \eqref{eq:LyapIneq} by solving it with respect to the
matrix $M$.
\end{rem}

The analysis is different when the unmatched perturbation
$\beta(t) \neq 0$. This can happen when attempting a robust
tracking of the hydraulic actuator in presence of the external
mechanical forces $F_L(t)$. This is discussed below.
\begin{lem}
Suppose that Assumption \ref{ass:1} is satisfied and there exists
$\beta_M>0$ such that $|\beta |\leq \beta_M$.  If
\eqref{eq:LyapIneq} is satisfied, then the trajectories of $e(t)$
satisfies the following bound:
\begin{equation}\label{eq:ultimatebound}
\Vert e(t)\Vert \leq {  \frac{2(\lambda_{max}(M))^{3/2}\beta_M}{\theta_V \mu(M)(\lambda_{min}(M))^{1/2} }  } \,,
\end{equation}
for positive constants $\mu(M)$ (dependent on $M$) and $ 0 <
\theta_V < 1$.
\end{lem}

\textit{Proof}. The proof is given in Appendix B.\newline
In Lemma 3, a free parameter $\theta_V$ is introduced in order to estimate the ultimate bound of the error \eqref{eq:ultimatebound}.  It is worth to be mentioned that $\theta_V$ is a variable meant to ensure the negative definiteness of the Lyapunov derivative of $V= e^TMe$ whenever \eqref{eq:ultimatebound} is satisfied. The bigger the $\theta_V$, the best the estimation \eqref{eq:ultimatebound}. However, it can be seen that as long as $\theta_V$ approaches to $1$, the negativeness of the Lyapunov derivative is compromised, and therefore the right hand side \eqref{eq:ultimatebound} tends to infinity. One can therefore establishes $\theta_V = \frac{1}{2}$ and then algorithmically make it smaller (conversely, greater) depending on the Lyapunov derivative and the estimated ultimate bound.

\subsection{STA-based sliding mode control design}

As shown in the previous section, if the variable $\eta(t)$, as in
\eqref{eq:v}, emulates the virtual, then the control objective for
mechanical sub-system will be achieved. In order to accomplish
this task, let us consider the sliding variable
\begin{equation}\label{eq:surface}
s(e) = \eta - \int v(e_1,\eta)dt +\kappa e_1 + \frac{4\tau E
A}{V_t} e_1
\end{equation}
such that if $s(e) =0$ then $\eta = \int v(e_1,\eta)dt -\kappa e_1
-\frac{4\tau E A}{V_t}e_1$.  Its dynamics results to
\begin{equation}
\begin{aligned}
\dot{{s} }(e) &= \dot{\eta}-  v(e_1,\eta) + \kappa \dot{e}_1 + \frac{4\tau E A}{V_t} \dot{e}_1 \\
&= \gamma_1 e_1 +  \kappa e_2
+\left(\gamma_2 - \frac{4 E C_{qp} }{V_t} \right)\eta  \\ &+ \frac{4\tau E C_{q} }{V_t}u(t) +
\bar{\delta}_3(t).
\end{aligned}
\end{equation}
Before introducing the proposed control law, let us prove the reachability of such sliding surface in the following Lemma.
\begin{lem}\label{lem:reach}
Assume that there exist a positive value $C_{e_2}$ such that $\vert e_2(t)\vert \leq C_{e_2}$.  Then, there always exist a control law (that does not depend on the perturbations) so that the trajectories of system \eqref{eq:errordynamics} converge to the manifold $s(t) \equiv 0$ in a finite time $T_s$, and are kept in such manifold for all $t\geq T_s$.
\end{lem}
\textit{Proof}. The proof is given in Appendix C.\newline
Let us now introduce the STA-based control law of the form:
\begin{equation}
\begin{aligned}
u(t) =& -\frac{V_t}{4\tau E C_{q} } \left[ k_1\rho \lceil s\rfloor^{1/2}
+\gamma_1 e_1 + \left( \gamma_2 \frac{4 E C_q}{V_t}\right)\eta
\right] \\ &- \frac{V_t}{4\tau E C_{q} }\left( \int k_2 \rho^2
\lceil s\rfloor^0 dt \right)\,,
\end{aligned}
\end{equation}
with the positive constants $k_1$, $k_2$ selected so as to obtain
the desired behavior of STA, and $\rho$ to be a scaling factor for
dealing with perturbations. It is important to notice that the
discontinuous part of control is responsible for robustness
properties of the STA against the perturbations whose derivative
is bounded. Moreover, discontinuity in the control signal is
avoided since such term is integrated. Therefore, in order to
analyze conditions posed on the perturbations, and the stability
of the control algorithm, it is necessary to investigate also the
dynamics of integral part. Setting a new variable \[ z =-
\frac{V_t}{4\tau E C_{q} }\left( \int k_2 \rho^2 \lceil s\rfloor^0
dt \right) + \bar{\delta}_3(t) + \kappa e_2\] leads to
\begin{equation}
\begin{aligned}
\dot{{s} }(e) &= -\rho k_1 \lceil s\rfloor^{1/2}+ z\,, \\
\dot{z}&=-\rho^2 k_2\lceil s\rfloor^0+ \dot{\bar{\delta}}_3(t) + \kappa \dot{e}_2  \,,
\end{aligned}
\end{equation}
where the whole perturbation in the second line is $\delta_z =
\dot{\bar{\delta}}_3(t) + \kappa \dot{e}_2$. Note that an exact
convergence of the variable $z$ to zero implies the theoretically
exact compensation of perturbations. This is because in order to
force $z$ to converge, it is required that the integral term of
the control law has theoretically exactly the same value as the
perturbation $\delta_z$. Then, the system becomes a standard STA
form
\begin{equation}\label{eq:stacl}
\begin{aligned}
\dot{{s} }(e) &= -k_1\rho \lceil s\rfloor^{1/2} + z\,,  \\
\dot{z}&=-k_2\rho^2\lceil s\rfloor^0 + \delta_z(t) \,.
\end{aligned}
\end{equation}

\begin{assum} \label{ass:2}
For all $t \in \mathbb{R}_+$, there exists a non-negative constant
$L$ such that
\begin{equation}
\begin{aligned}
\vert \delta_z(t,e)\vert &\leq L\,.
\end{aligned}
\end{equation}
\end{assum}

\begin{lem}\label{lemma:sta}
Given the Assumption \ref{ass:2} is satisfied for the sub-system
\eqref{eq:stacl}. Design the $k_1$, $k_2$ gains such that
\begin{equation}
A_{k}= \begin{bmatrix}
-k_1 & 1 \\ -k_2 & 0
\end{bmatrix}
\end{equation}
is a Hurwitz matrix.  Let $M_k$ be the solution of $A^T_kM_k +
M_kA_k = -\mathbb{I} $, if the scaling factor satisfies
\begin{equation}\label{eq:rho}
\rho > {2 L \lambda_{max}(M_k) }\,.
\end{equation}
Then, the trajectories of the dynamic system \eqref{eq:stacl}
converges to the origin in a finite time.
\end{lem}
{\textit{Proof}}. The proof is given in Appendix D.

\begin{rem}
Note that with this approach, one can set the gains $k_1$, $k_2$
so as to obtain a certain desired behavior, which can then be
scaled by the variable $\rho$ as in \cite{Seeber2017},
\cite{Levant1993}, and \cite{Moreno2012}. One option is to set the
gains such that it optimizes a certain criteria in the sense of
\cite{Ventura2019}.
\end{rem}

\section{Control synthesis}\label{sec:design}

\subsection{Surface Design}

\begin{figure}[h]
\centering
  \includegraphics[width=0.75\linewidth]{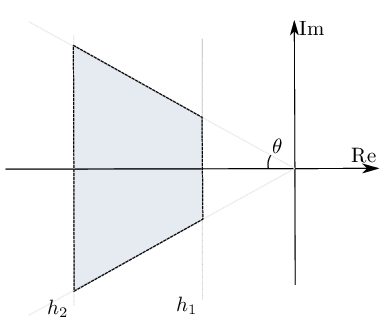}
  \caption{Desired LMI region in the complex plain.}
  \label{fig:lmi}
\end{figure}

For the surface design, recall the Lyapunov-type inequality
\begin{equation}
MA_n+A_n^TM -K^TB^TM - MBK+\Psi H^TM + \Psi MH<0
\end{equation}
and consider the change of variable $Y = M^{-1}$ such that
multiplying with $Y$ both sides of the above expression yields
\begin{equation}
Y(A_n+ \Psi H)^T   +(A_n+ \Psi H)Y + YK^TB^T + BKY<0\,.
\end{equation}
Afterwards, setting the new variable to be found as $N = KY$ such
that one can compute $K = NM$, the inequality becomes
\begin{equation}
Y(A_n+ \Psi H)^T   +(A_n+ \Psi H)Y +N^TB^T + BN< 0\,.
\end{equation}
Additionally, let us consider the following constraints on the LMI
regions to be computed as
\begin{equation*}
\begin{gathered}
Y\mathcal{W}^T(A_n,H)  + \mathcal{W}(A_n,H)Y + \mathcal{N}(N,B) +2h_1Y <0 \\
Y\mathcal{W}^T(A_n,H)   + \mathcal{W}(A_n,H)Y+ \mathcal{N}(N,B) +2h_2Y >0 \\
\begin{bmatrix}
\sin(\theta)C_{11} & \cos(\theta)C_{12} \\ \cos(\theta)C^T_{12}& \sin(\theta)C_{22}
\end{bmatrix}<0,
\end{gathered}
\end{equation*}
with $\mathcal{W}(A_n,H) = A_n + \Psi H$, $\mathcal{N}(N,B) = BN + N^TB^T$, and
\begin{align*}
C_{11} &= \left[ YW^T(A_n,H)   +\mathcal{W}(A_n,H)Y + \mathcal{N}(N,B)  \right] \\
C_{12} &= \left[YA_n^T-N^TB^T - A_nY +BN\right]\\
C_{22} &= \left[ Y\mathcal{W}^T(A_n,H)   \mathcal{W}(A_n,H)Y + \mathcal{N}(N,B)\right] \,.
\end{align*}
The above formulated constraints generate a region in the complex
plane as schematically depicted in Fig. \ref{fig:lmi}.
The next to be analyzed is selection of the constants $h_1$, $h_2$
and $\theta$ for the LMI-based surface design.

\subsubsection{Assignment of $h_2$:}\label{ssec:h_2}

In order to assign a proper boundary value $h_2$, one can consider
an approximated first-order response of a linear system of the
form $c_{h_{2}}(t) = 1 - \exp(t/T_{h_{2}})$. In order to obtain a
time constant such that the first-order response to the step as
$\lim_{t\rightarrow\infty}  c_{h_2}(t) = \bar{y}_2$, the response
should reach $63.2\%$ of the total desired amplitude. Then, from a
straightforward geometrical reasoning, the associated time
constant can be approximated by
\begin{equation}
 T_{h_2} \approx \sqrt{(0.632y_d)^2 - (\bar{y_2})^2}\,.
\end{equation}
and consequently $h_2 = 1/T_{h_2}$.

\subsubsection{Assignment of $h_1$:} \label{ssec:h_1}

Consider the mechanical sub-system and assume that the pressure
difference between the chambers is the given input. Then, in terms
of the pressure to control the second-order dynamics
\begin{equation}
\frac{d}{dt}\begin{bmatrix}
x_1 \\ x_2
\end{bmatrix} = \begin{bmatrix}
0 & 1 \\ 0 & -\frac{\sigma}{m}
\end{bmatrix}\begin{bmatrix}
x_1 \\ x_2
\end{bmatrix}+ \begin{bmatrix}
0 \\ \frac{1}{\tau m}
\end{bmatrix}A P\,,
\end{equation}
and assuming $ AP = -k_hx_1 -\sigma_h x_2$ with the feedback
constants in a hypothetical closed-loop
\begin{equation}
\frac{d}{dt}\begin{bmatrix}
x_1 \\ x_2
\end{bmatrix} = \begin{bmatrix}
0 & 1 \\ -\frac{k_h}{\tau m}& -\frac{\sigma\tau + \sigma_h}{\tau m}
\end{bmatrix}\begin{bmatrix}
x_1 \\ x_2
\end{bmatrix},
\end{equation}
or its summarized compact form $ \frac{d}{dt}\begin{bmatrix} x_1 &
x_2 \end{bmatrix}^T = A_h\begin{bmatrix} x_1 & x_2
\end{bmatrix}^T$, one can compute the eigenvalues as
\begin{equation}
\lambda_{1,2}(A_{h}) = -\frac{\sigma \tau + \sigma_h}{2\tau m}\pm \frac{1}{2\tau m}\sqrt[2]{\left(\tau \sigma + \sigma_h \right)^2 - 4k_h \tau m}\,.
\end{equation}
The critically damped case is when $(\tau \sigma + \sigma_h )^2 -
4k_h \tau m = 0$, and both eigenvalues have the same real part.
Then, the step response of the position as output has a form
$x_1(t) = 1 - \exp^{-\omega_n t}(1+\omega_n t)$ where $\omega_n
=\frac{\sigma \tau + \sigma_h}{2\tau m}$. At the same time, if one
approximates the step response by a first-order behavior, one can
compute the time constant as
\begin{equation}
T_{h_1} = \frac{2m\tau}{\sigma\tau+\sigma_h}\,,
\end{equation}
which implies that for a desired $T_{h_1}$, one can determine
\begin{equation}
\sigma_h = \frac{2m\tau}{T_{h_1}} - \sigma\tau \,.
\end{equation}
Then, for a desired behavior of the mechanical sub-system, being
related to its time constant $T_{h_1}$, one obtains
\begin{equation}
h_1 =  -\frac{\sigma \tau + \sigma_h}{2\tau m}
\end{equation}

\subsubsection{Parameter $\theta$:}\label{ssec:theta}
For many practical reasons,  it is convenient to have only real negative eigenvalues such that the mechanical response does exhibit an oscillating response. This can be done if the parameter $\theta = 0$, and therefore the cone-constrained LMI region collapses only into the real line.  However,  this can cause computational problems or the feasibility of the LMI's may be compromise.  Parameter $\theta$ is introduced such that it can be adjusted in order to find a solution to the LMI,  by choosing it as small as possible.

\subsubsection{Bound for $\psi(t,e)$:}\label{ssec:psi}

Moreover, assume that the largest mechanical perturbation is due
to nonlinear fiction term, cf. \eqref{eq:friction},
and rewrite it as
\begin{equation}
\psi(t,x_2) = \frac{ \tanh( \vartheta x_2)\left[ F_c +
(F_s-F_c)\text{exp}\left(-\vert x_2 \vert^{\phi} \chi^{-\phi}
\right) \right] }{x_2}\,,
\end{equation}
which has its maximum as $x_2$ tends to zero. Then, one can find
the bound as
\begin{equation}
\Psi = \sup \lim_{x_2\rightarrow 0} \psi(t,x_2)\,.
\end{equation}

\subsection{STA design}

In order to compute the $L$ value, take the perturbation
$\delta_z$ and calculate its bound as
\begin{equation}
\begin{aligned}
\vert \delta_z \vert &\leq  \vert
\bar{\delta}_3(t)\vert + \kappa\vert \dot{e}_2(t)\vert\,.
\end{aligned}
\end{equation}
Assuming that the system will be in a compact set, there exists a
constant positive value $\overline{ \ddot{q} }$ such that $\vert
\dot{e}_2\vert \leq \overline{ \ddot{q} }$. This way, the value of
$L$ can be determined by
\begin{equation}
L =  L_3 + \kappa \overline{ \ddot{ q} }
\end{equation}

\section{Comparison with other SMC strategies}\label{sec:comparison}

For the sake of clearness in the comparison presented in this section,  the approach proposed in this paper will be called IS-STA.

A qualitative comparison with respect to the approach in \cite{oliveira2018} will be performed, taking into account that both of the approaches are based on the STA algorithm. Additionally, for both methodologies the velocity measurement is not needed in order to apply the control law.

The approach of \cite{oliveira2018} consists in three main steps: first, a tracking reference model is designed, secondly, a variable gain Levant's differentiator is implemented, and finally, a variable gain STA based control law is designed.  The variable gains for the differentiator and STA are updated using a norm observer. The sliding surface designed depends on the output and its derivatives, which are estimated by the variable gain Levant's differentiator. It is important to remark, that the parameters used \cite{oliveira2018} are tuned so that the tracking performance for both of the approaches are similar.

  According to \cite{oliveira2018}, the following reference model is proposed:
\begin{equation}
y_m = \frac{(25 )^2}{(s+25)^2}r\,.
\end{equation}
The chosen Input to Output state variable filters have the form:
\begin{equation}
w_u(s) = \frac{1}{(s + 5)^2}u\,, \quad w_y(s) = \frac{1}{(s+5)^2}q\,.
\end{equation}
The variable gain differentiator implemented is the following
\begin{equation}
\begin{aligned}
\dot{\hat{e}}_1 &= -3L_{vgst}^{1/4}(t,\hat{x}) \lceil \hat{e}_1 - e_1\rfloor^{3/4} -2(\hat{e}_1 -e_1) + \hat{e}_2\,, \\
\dot{\hat{e}}_2 &= -2.5L_{vgst}^{1/3}(t,\hat{x})\lceil \hat{e}_1 - e_1\rfloor^{2/3} -3(\hat{e}_1 -e_1) + \hat{e}_3\,,\\
\dot{\hat{e}}_3 &= -1.5L_{vgst}^{1/2}(t,\hat{x})\lceil \hat{e}_1 - e_1\rfloor^{1/2} -2(\hat{e}_1 -e_1) + \hat{e}_4\,,\\
\dot{\hat{e}}_4 &= -1.1L_{vgst}(t,\hat{x})\lceil \hat{e}_1 - e_1\rfloor^0 -(\hat{e}_1 -e_1) \,.
\end{aligned}
\end{equation}
where the gains comes from:
\begin{equation}
\begin{aligned}
\dot{\hat{x}} &= -0.8 \hat{x}(t) + 10 + 1.5\left\Vert \begin{bmatrix}
w_u & w_y
\end{bmatrix}^T\right\Vert\,,\\
L_{vgst}(t,\hat{x}) &= 1.2\vert \hat{x}\vert + 2\vert u\vert + 7\,.
\end{aligned}
\end{equation}
The control strategy consists of the following sliding surface and controller
\begin{equation}
\begin{aligned}
\hat{\sigma}(t) &= \hat{e}_3 + 2(25)\hat{e}_2 + (25)^2e_1\\
u_{vgsta} &= -\tilde{k}_1(t,\hat{x})\phi_1(\hat{\sigma}) - \int_0^t \tilde{k}_2(\tau,\hat{x})\phi_2(\hat{x})d \tau\,,\\
\phi_1(\hat{\sigma}) &=  \lceil \hat{\sigma}\rfloor^{1/2} + \hat{\sigma}\,, \\
\phi_2(\hat{\sigma}) &= \frac{1}{2}\lceil \hat{\sigma}\rfloor^0 + \frac{3}{2}\lceil \hat{\sigma}\rfloor^{1/2} + \hat{\sigma}\,,
\end{aligned}
\end{equation}
where the gains $\tilde{k}_1(t,\hat{x})$ and $\tilde{k}_2(t,\hat{x})$ are designed according to \cite{oliveira2018}, taking into account the following bounds of the perturbations $\rho_1 = 0$ and $\rho_2 = 10\vert \hat{e}_2\vert + 5\vert e_1\vert + \hat{x} + 1$ and the constants $\delta_k = 0.01$, $\varepsilon_k = 1\times 10^{-3}$,  resulting in:
\begin{align*}
\tilde{k}_1 &= \delta_k + \frac{1}{4\varepsilon_k}\rho_2^2 + 2\varepsilon_k\rho_2 + \varepsilon_k + 2\varepsilon_k\left(1 + 4\varepsilon_k^2\right)\,,\\
\tilde{k}_2 &= 1 + 4\varepsilon_k^2 +2\varepsilon_k \tilde{k}_1\,.
\end{align*}
The simulations were performed using a band limited white noise in the measurements, similar to the one presented in the real plant sensors, with noise power of $1\times 10^{-9}$ and a sample time of $5\times 10^{-6}$.  The details about the designed trajectories are presented in the experimental evaluation.

\begin{figure}[h]
\centering
 \includegraphics[width=1.1\linewidth]{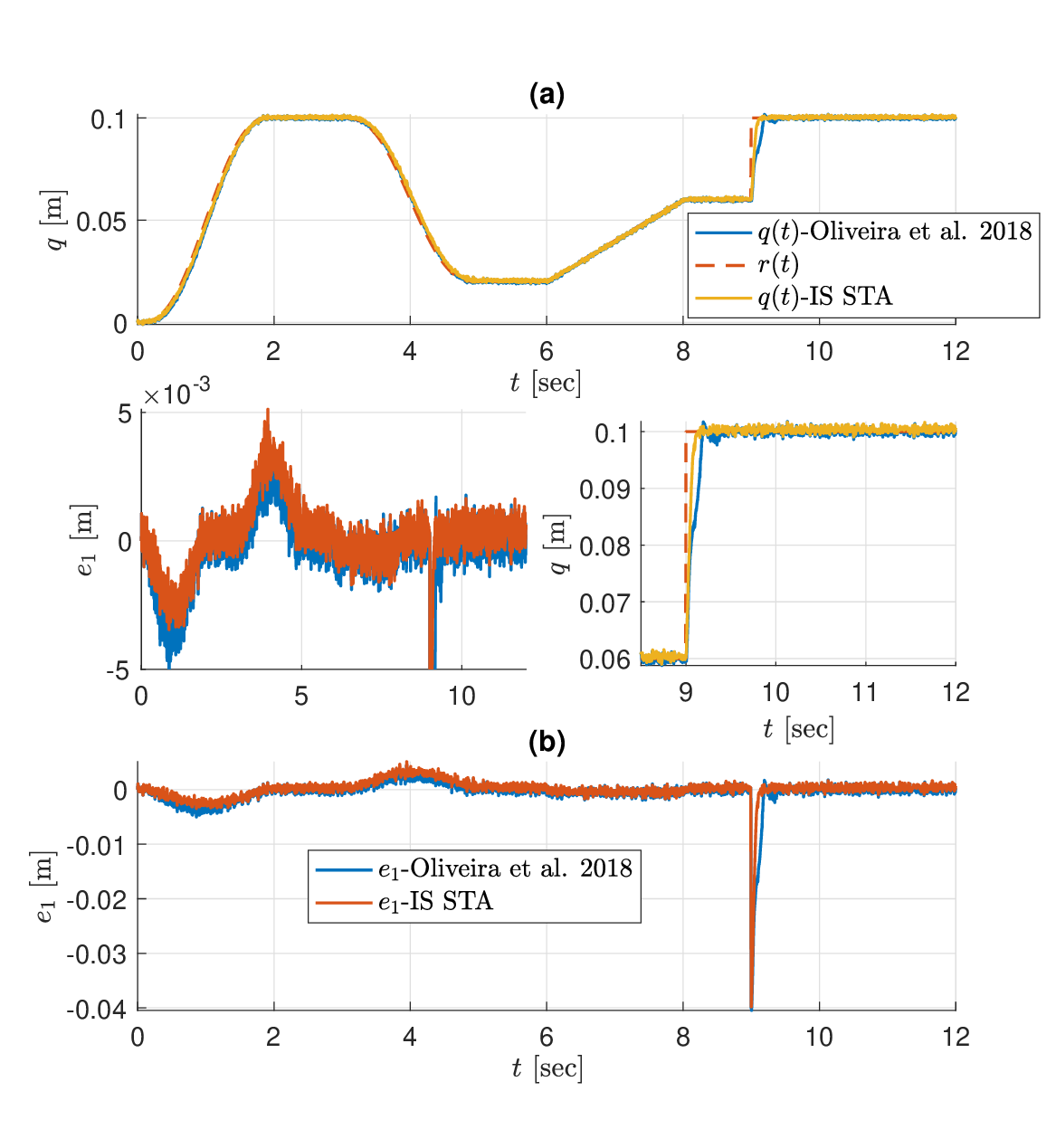}
  \caption{Comparison of the position $q$ and tracking error $e_1$.  $r(t)$ corresponds to the reference to be tracked, $q(t)$- IS STA is the proposed approach and $q(t)$- Oliveira et al. 2018 is the approach presented in \cite{oliveira2018}.}
  \label{fig:qcompnoisy}
\end{figure}

It can be observed in Figure \ref{fig:qcompnoisy}, that the tracking is well performed by the two methodologies.  In Figure \ref{fig:qcompnoisy}-(b) the errors are presented. Although the IS STA appears to be better in the first $3$ seconds,  the approach of \cite{oliveira2018} works similarly well in second stage of the trajectory, from $t = 3$ to $t = 5$.  During the transient of the step response, the IS STA slightly outperforms \cite{oliveira2018}.

\begin{figure}[h]
 \includegraphics[width=1.1\linewidth]{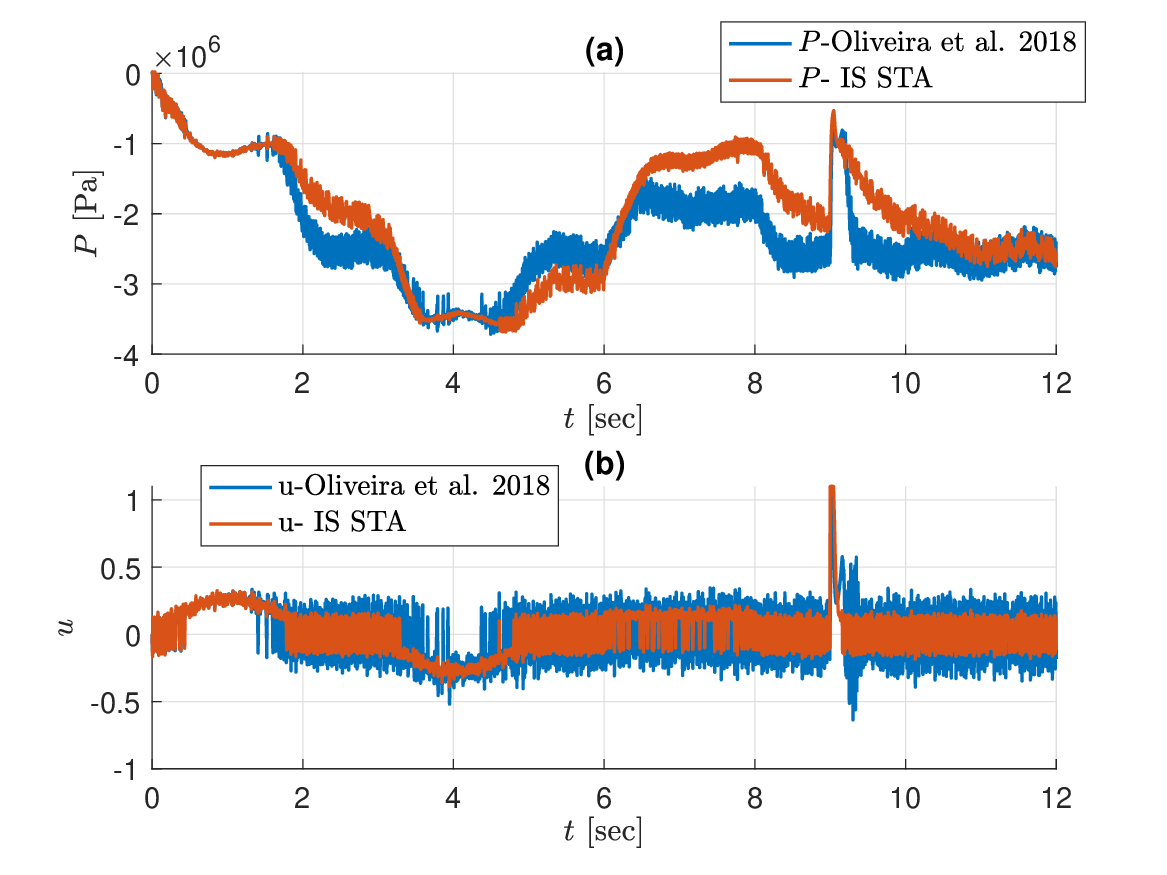}
  \caption{Comparison of the pressure variable $P$ and control signal $u$ corresponds to the reference to be tracked,  IS STA is the proposed approach and Oliveira et al. 2018 is the approach presented in \cite{oliveira2018}.}
  \label{fig:Pcompnoisy}
\end{figure}

In Figure \ref{fig:Pcompnoisy} the pressure diference $P$ and the control signal $u(t)$ are presented.  From the control signal $u(t)$ it is clear that $u$-IS STA is smaller in amplitude than the one from the $u$-Oliveira et al. 2018. This can be caused by the noise in the measurement which is propagated not only in the Variable Gain Super-Twisting, but also via the third order differentiator.  One may decrease gains in order to reduce the effect of the noise in the control input. However, this also make the performance to decrease.

\section{Experimental Evaluation}\label{sec:numeval}
\subsection{Experimental setup}

\begin{figure}[h]
\centering
  \includegraphics[width=0.6\linewidth]{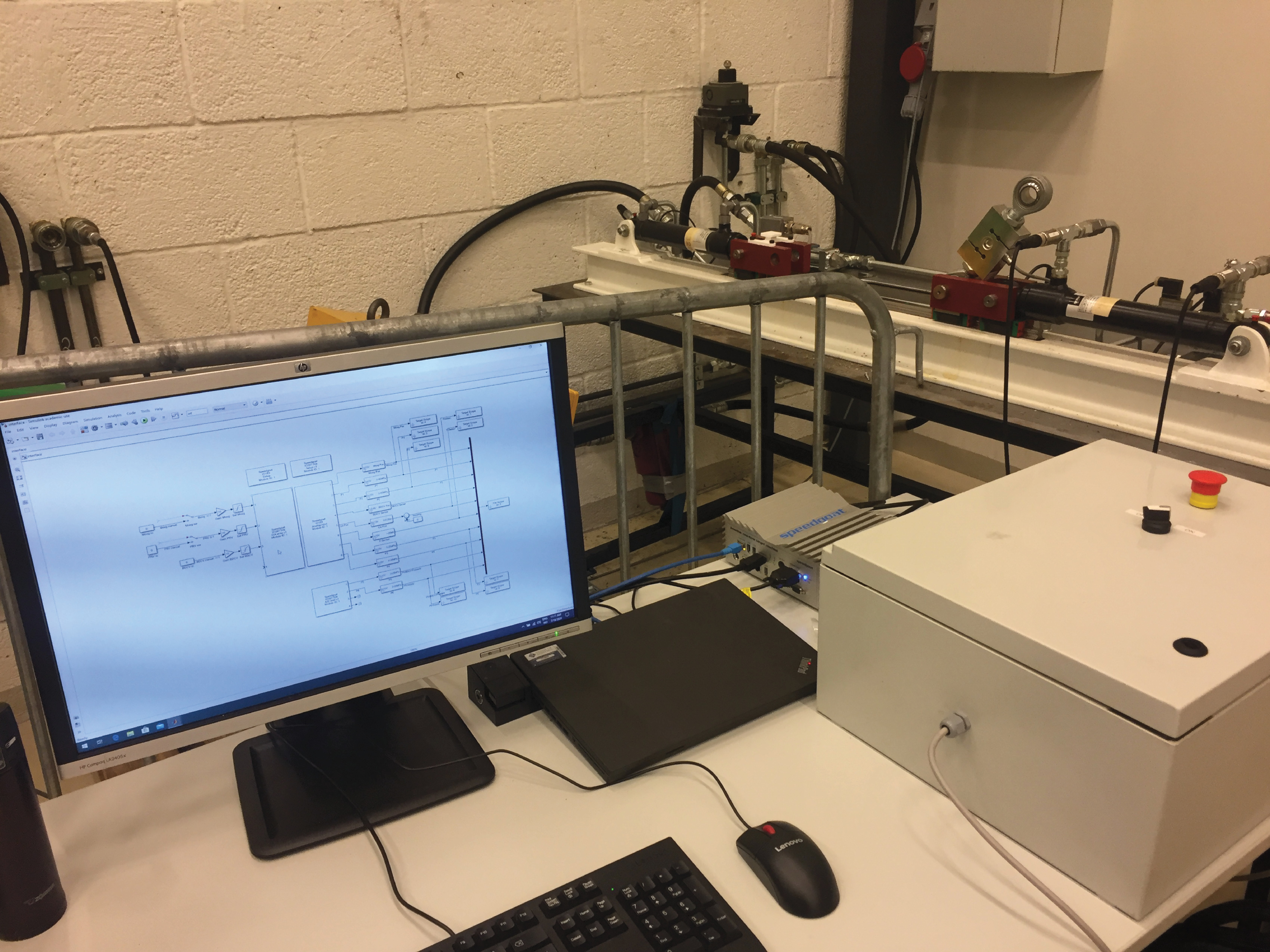}
  \caption{Experimental test bench of hydraulic actuator (laboratory view).}
  \label{fig:setup}
\end{figure}

\begin{figure}[h]
\centering
  \includegraphics[width=0.6\linewidth]{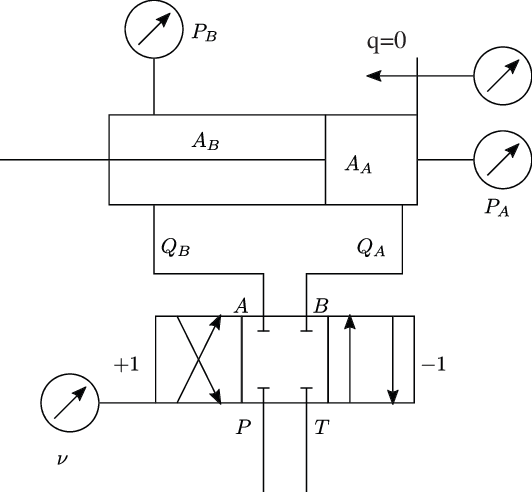}
  \caption{Schematic diagram of the hydraulic circuit of the actuator system.}
  \label{fig:hydrcircuit}
\end{figure}

In this section, the experimental validation of the proposed
control approach is provided. The presented control solution is
implemented in the laboratory test bench \cite{Pasolli2018}, see
laboratory view shown in Fig. \ref{fig:setup}. The hydraulic
setup is composed by a one-side-rod cylinder with a moving load
carrier, mounted in a linear slide bearing, and one translational
degree of freedom $q$.
The cylinder is connected via a 4 ways / 3 positions servo-valve
with an internal closed-loop spool control, connected to the
hydraulic power unit, cf. schematic diagram of the hydraulic
circuit shown if Fig. \ref{fig:hydrcircuit}.
The supply parameters are set to $10$ MPa pressure and $40$
liter/min volumetric flow. More details on the hardware components
can be found in \cite{Pasolli2018}. Since the low-level controlled
servo-valve receives the input reference signal in percentage of
the orifice opening, the control signal $u$ and correspondingly
$g$ (cf. with section 6.3) is inherently unitless i.e. $u \in
[-1,+1]$. Worth noting is also that the control algorithms are
implemented in Simulink\textregistered Real-Time on a
Speedgoat\texttrademark platform with 2 kHz sampling rate.

\subsection{Motion trajectories}

The trajectories were designed as piece-wise continuous functions
for different time instants. From the time $t=0$ to $t=2$, a
quintic smooth polynomial trajectory is tracked, starting from
$q(0) = 0$ m to $q(2) = 0.1$ m in a fashion of smooth robotic
trajectories, cf. \cite{Spong2006}. Then, the value is maintained
until $t = 3$ s and then switched to another quintic polynomial
trajectory from $q(3) = 0.1$ m to $q(5) = 0.02$ m. Then a ramp is
assumed from $t = 6$ s to $t=8$ s and, finally, a step from $q(9)
= 0.1$ m is imposed. Worth noting is that although the final step
is not a smooth trajectory, as required for $r(t)$, the proposed
control approach is able to track it as well.

\subsection{Implemented control scheme}

The first implementation issue of the applied control scheme is
the dead-zone effect in the servo-valve opening orifice. This
effect, as shown in \cite{Pasolli2018}, is (nearly) static around
the $10\, \%$ of the opening range. Therefore, the dead-zone is
compensated in feed-forwarding via an inverse map of the form:
\begin{equation}
\tilde{u} = \frac{1}{2}D_s\text{sign}(u)+ u
\end{equation}
where $D_s>0$ is the dead-zone size.

\begin{figure}[h]
\centering
  \includegraphics[width=1.1\linewidth]{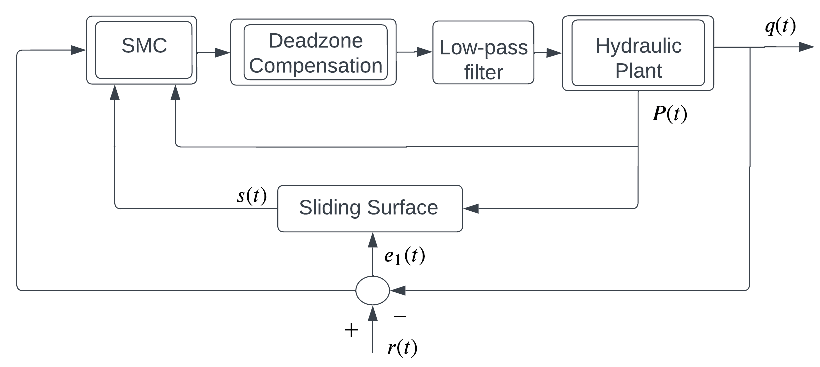}
  \caption{Schematic of the overall control strategy implemented.}
  \label{fig:OverallControlStrategy}
\end{figure}

Second, in the control related modeling and analysis, the servo
valve dynamics was not taken into account. Such actuator dynamics
can be pre-compensated in the same way as dead-zone effect. From
the previous work \cite{Pasolli2018}, the parameters of the
servo-valve input-output behavior of the form \eqref{eq:valve} are
obtained. Then, a low-pass filter is implemented in order to count
for the fast (but still not zero) dynamics of the servo valve. The
overall control value has then the form
\begin{equation}\label{eq:finalcontrol}
g = \frac{1}{\left(\mu_c \varphi + 1\right)^2}\left[
\frac{1}{2}D_s\text{sign}(u)+ u  \right]
\end{equation}
where $\varphi$ is the Laplace variable and $\mu_c>0$ is a design
parameter chosen to be equal to the servo-valve actuator time
constant. The identified parameters are given in
\cite{Pasolli2018}. Moreover, we stress that due to the dead-zone
compensation the chattering effect may be increased in the control
zero crossing. In that sense, the set low-pass filter complies
along with the servo-valve bandwidth, reducing the chattering of
the effective control value $g(t)$. The overall control strategy implemented is presented in Figure \ref{fig:OverallControlStrategy}.

\subsection{Auxiliary frequency domain analyze}

Based on the experimental evaluations, the authors notice some
practical issues that are worth to be addressed as below. The
stability of the sliding dynamics can also be lost by changing the
design parameters $h_1$ and $h_2$. In this section, based on the
concept of \textit{phase deficit} \cite{Boiko2011}, a possible
explanation to this phenomena is given. In a nutshell, the phase
deficit represents a certain measure of robustness for a
finite-time controller, similar to concept of the phase margin
usable for asymptotic linear controllers. It can be seen as the
angular difference of the reciprocal of negative inverse
Describing Function (DF) of a nonlinear control law and the
plant's Nyquist plot \cite{Boiko2011}.

Let us analyze the output of the system that is assumed to have
the form: $y(t) = a_y\sin(\Omega t + \phi_y)$, where $\Omega$ is
the frequency and $a_y$ is the amplitude of steady oscillations,
and $\phi_y$ is the phase lag. Assuming that the dynamic sliding
surface is a periodic input $s(t) = a_s \sin(\omega t)$ being $a_s>0$ the amplitude of the oscillations in the surface $s$, it is
possible to find the oscillatory reaction $y(t)$, and more
important, an explanation on how it is affected by the design of
the variable $h_2$. Consider that the transfer function function
of the closed loop has the form
\begin{equation} G_f(\varphi) = \frac{1}{\frac{1}{T_s}\varphi
+ N_f(a_y)}\,
\end{equation}
where the positive constant $T_s$ is a time constant dependent on
the design parameters $h_1$ and $h_2$, and $N_f$ is the DF of the
STA
\begin{equation}
N_f(a_y) = \frac{2k_1L\gamma(1.25)}{  \sqrt{\pi} \gamma(1.75)
}a_y^{-1/2} = \gamma_a a_y^{-1/2}\,,
\end{equation}
where $\gamma(\cdot)$ is the gamma-function, as it is presented in
\cite{Boiko2020}. Setting $a_1 = \frac{4k_2L^2}{\pi \omega} $ and
following \cite{Boiko2020}, one can compute the amplitude of the
propagated oscillations as: $a_y = a_1\vert G_f(jw)\vert$, or
equivalently:
\begin{equation}
\frac{w^2}{T_s^2}a_y^2 + \gamma^2a_y - a_1^2 = 0\,,
\end{equation}
whose solution is given by
\begin{equation}
a_y =  \frac{T_s^2}{2w^2}\left(-\gamma^2 + \sqrt{\gamma^4 + 4T_s^2\omega^2a_1^2} \right)\,.
\end{equation}
Computing its \textit{phase deficit} it yields \cite{Boiko2020} :
\begin{equation}
\phi_d = \frac{\pi}{2} - \arctan\left( \frac{1}{\sqrt{2}} \sqrt{-1 + \sqrt{1 + \frac{64k_2^2\rho^2}{T_s^4 \pi^2 \gamma^4} }  } \right)\,,
\end{equation}
ensuring that the \textit{phase deficit} approaches to zero as
$T_s$ tends to zero as well. This gives us an explanation why the
nominal system's response can not be made arbitrarily fast without
compromising the system's stability, and implies the selection of
$h_2$ margin should be additionally restricted, cf. with section
5.1.1.

\subsection{Experimental results}

For the control evaluation, the computed parameters for the LMI
conditions are $h_2= -5$ and $h_1 = -1$, these from the assigned
time constants $T_{h_2}= 1/5$ s and $T_{h_1} = 1$ s. The parameter
$\theta$ is left to be small enough such that the LMI is still
feasible but, at the same time, tries to restrict the imaginary
part of the resulting poles. In this work, $\theta = \pi/20$ was
chosen. From the aforementioned analysis one can get $\Psi = 0.5$.
Finally, for the STA, we assume the gains which minimize the
amplitude of chattering as  $k_1 = 1.1$ and $k_2= 2.028$, see
\cite{Ventura2019}, so that the scaling condition becomes $\rho
> 2(1.6303)L$. The computed boundary constant is $L=1.347$, and the
assigned scaling factor is $\rho = 10$. A summary of the parameters used for this experimental results is presented in table \ref{tab:table_parameters} with the corresponding methodology section.

\begin{table}[h!]
  \begin{center}
     \caption{Summary of the parameters and its corresponding section.}
    \label{tab:table_parameters}
    \begin{tabular}{l|c|r} 
      \hline\hline
      \textbf{Parameter:} & \textbf{Value:} & \textbf{Design section:}\\
      \hline\hline
      $h_1$ & 1110.1 & Section \ref{ssec:h_1} \\ \hline
      $h_2$ & 10.1 & Section \ref{ssec:h_2} \\ \hline
      $\theta$ & 23.113231 & Section \ref{ssec:theta} \\ \hline
      $k_1$ & 1.1  &   c.f. \cite{Ventura2019} \\
      $k_2$ & 2.028 &   \\ \hline
      $\rho$ & 10 & Section \ref{ssec:psi} \\ \hline
    \end{tabular}
  \end{center}
\end{table}

\begin{figure*}[h!]
\centering
  \includegraphics[width=1\linewidth]{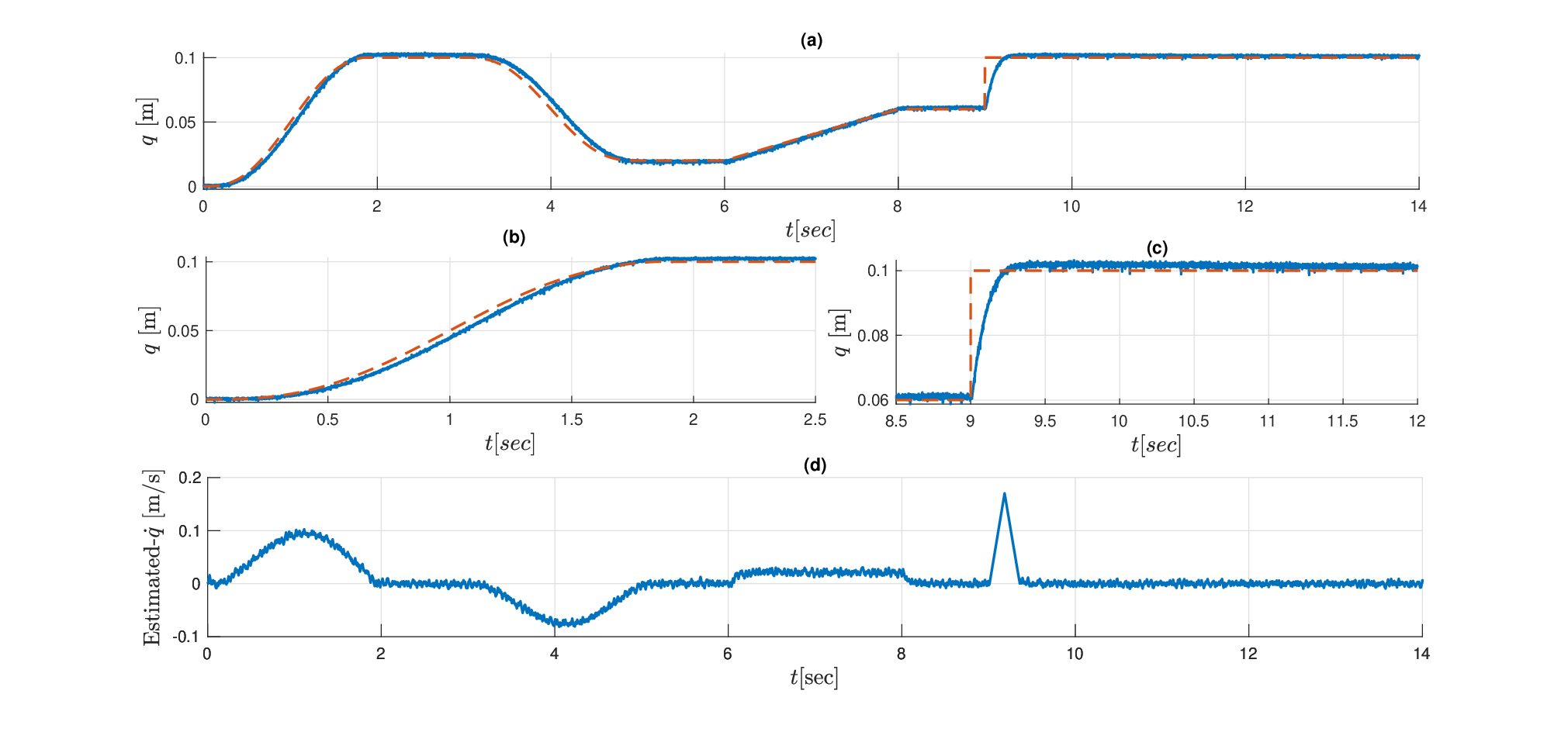}
  \caption{Measured position (versus reference) in (a), zoom from $t = 0$ to $t = 2.5$ seconds is presented in (b),  zoom on the tracking position from $t = 8.5$ to $t = 12$ seconds is presented in (c) and the estimated velocity of the controlled motion trajectory is shown in (d).}
  \label{fig:q}
\end{figure*}

\begin{figure*}[h!]
\centering
  \includegraphics[width=1\linewidth]{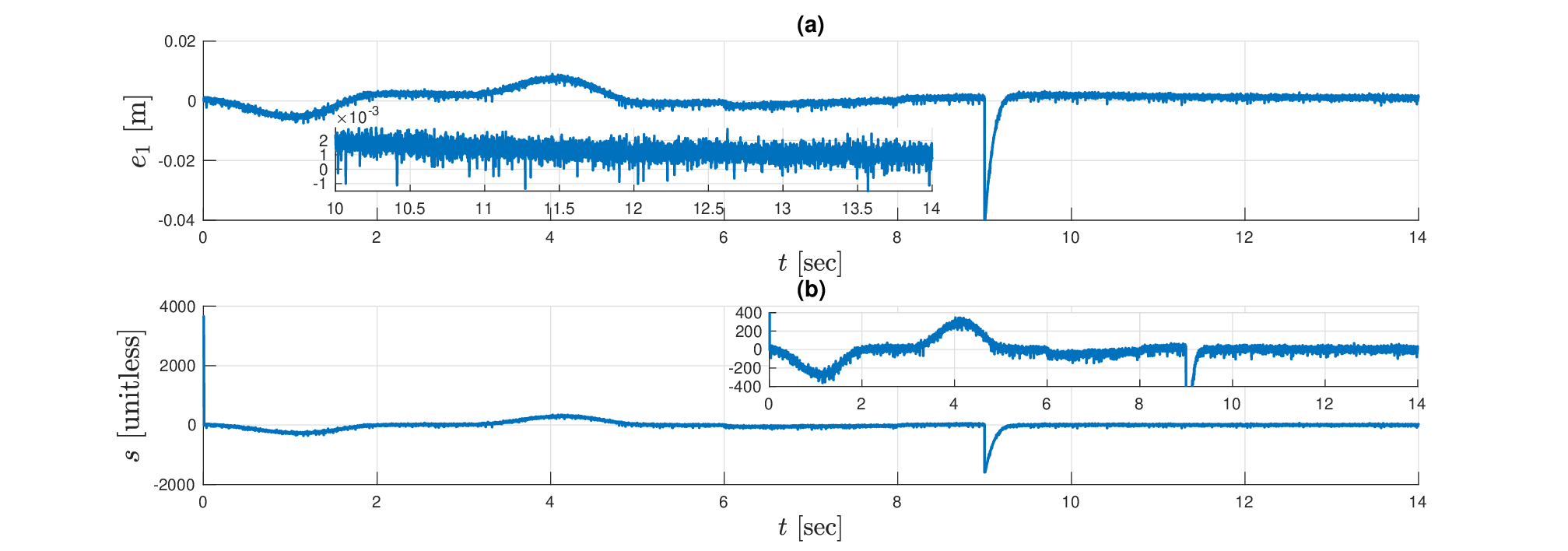}
  \caption{Measured output tracking error in (a) and the sliding mode variable in (b).}
  \label{fig:s}
\end{figure*}

The designed motion trajectory is evaluated experimentally, as
shown by the dotted line in Fig. \ref{fig:q} (a). It can be seen
that the cylinder position $q(t)$ follows the desired trajectory,
being uniformly close to the reference value $r(t)$. The measured
piston position is shown by blue solid line and the reference is
shown by the dotted black line in Fig. \ref{fig:q} (a). It can be
recognized that the position is tracked accurately even in the
case of a discontinuous (step) change at $t = 9\,\, \text{s}$. The
corresponding tracking error is also shown in Fig. \ref{fig:s}
(b). Although the velocity is not available and, thus, not used in
the proposed control approach, it is estimated using robust exact
differentiator toolbox \cite{Andritsch2021} for the sake of a
better visualization of the controlled motion. Figure \ref{fig:q}
(b) discloses that the estimated velocity is well in line with
derivative of the reference value.

\begin{figure*}[h!]
\centering
  \includegraphics[width=1\linewidth]{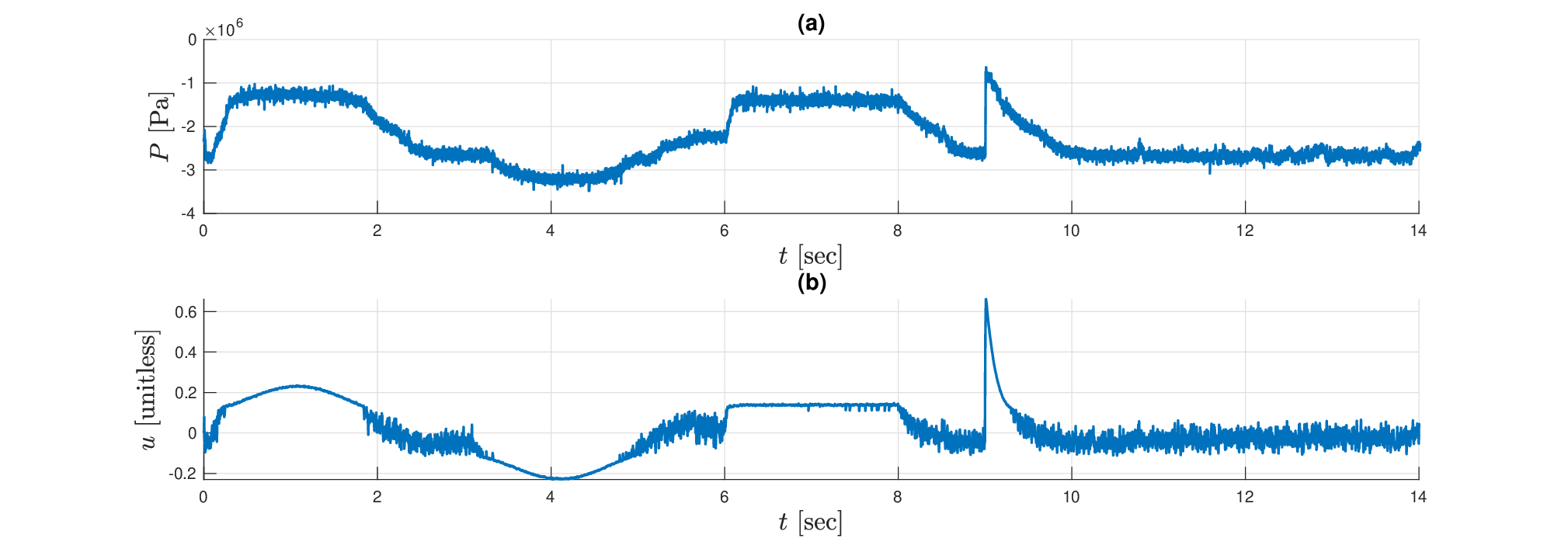}
  \caption{Measure differential pressure (as virtual control) in (a) and the control signal in (b).}
  \label{fig:p}
\end{figure*}

Figure \ref{fig:p} shows the differential pressure, i.e. measured
pressure difference between the right and left chambers, in (a)
and the resulted control signal which is applied to the
servo-valve in (b). Worth recognizing is that the control signal
is continuous. Some chattering pattern, where the control signal
appears to oscillate between $\pm\, 0.1$, appears at several
phases of steady-state and corresponds to the dead-zone effect in
the servo-valve. Recall that the latter is pre-compensated in
feed-forward by the dead-zone inverse. Also note that the
servo-valve blocks energizing the hydraulic circuits within the
dead-zone, so that a low-amplitude chattering is slightly outside
of $\pm\, 0.1$. Since the second-order actuator dynamics provides
the corresponding low-pass filtering, such dead-zone related
chattering is irrelevant for loading of the hydraulic circuits.
From Fig. \ref{fig:p} (a), the pressure is acting as virtual
control for the sub-mechanical system. It is important to recall
that the $P$ variable is the differential pressure so that its
negative values have reasoning for the selected motion direction.
Also worth recalling that within dead-zone, the valve is locking
the hydraulic circuit from one side and, consequently, one of the
chambers remains pressurized. This is explaining the non-zero
$P(t)$ values where there is no apparent motion of the cylinder
piston, cf. Fig. \ref{fig:q} and Fig. \ref{fig:p} (a). The load
pressure magnitude remain below $40\,\, \text{bar}$ while the
supply pressure is $100\,\, \text{bar}$. Figure \ref{fig:p} (b)
discloses also the continuous control signal (apart from the
dead-zone chattering) and that the control signal is never
saturated.
\newline

In Fig. \ref{fig:s}, zoom-in plot of the output tracking error and
the sliding variable are shown in (a) and (b), respectively.
Important to notice is that the sliding mode may be temporary lost
when the assumed conditions about perturbations, cf. (11) and
(13), are violated, like in case of the applied step reference at
$t=9$ s. Even then, the sliding mode is subsequently recovered and
the tracking is well performed.

\begin{figure}[h!]
\centering
  \includegraphics[width=1\linewidth]{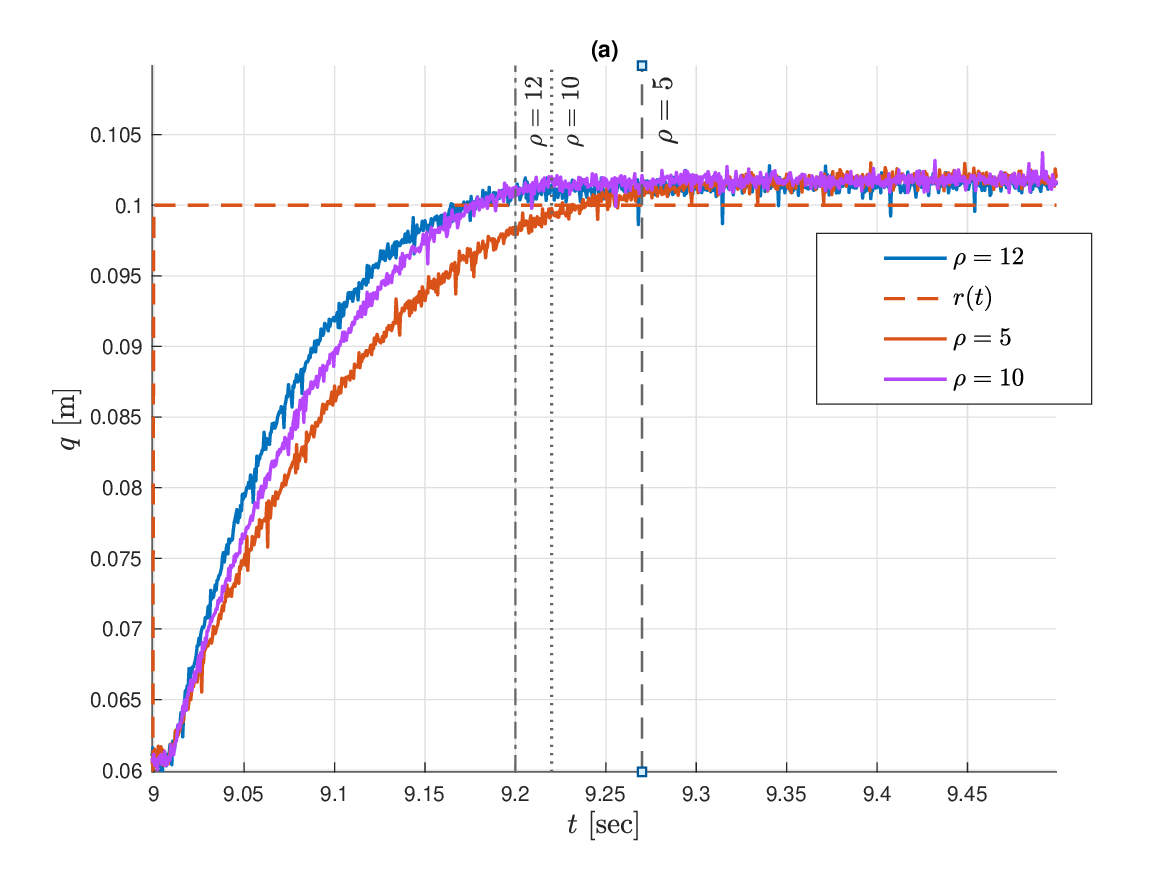}
  \caption{Comparison of the step response with different values of $\rho$.}
  \label{fig:steps}
\end{figure}

It is important to mention that the convergence velocity may be increased by the scaling factor $\rho$. In Figure \ref{fig:steps} the step response with respect to different values of $\rho$ is presented.  However, the bigger the gain $\rho$, the bigger the control input. Another important issue with increasing the gain is the amplitude of the chattering effect.  Increasing the gain may considerable increase the chattering produced by the noise.

A brief video of the presented experimental results is available
under \url{https://youtu.be/_dnwD7Ens-k}.

\subsection{Quantitative measures of performance}

In this section, some quantitative measures of the experimental results are given.  Let us define the following performance indices:
\begin{enumerate}
\item Maximal absolute value of the tracking error
\begin{equation}
M_e = \max_{i=1,2,\ldots,N} \left\lbrace \vert e_1(i)\vert \right\rbrace\,,
\end{equation}
being $N$ the number of obtained values recorded.
\item Average tracking error
\begin{equation}
\mu_e = \frac{1}{N}\sum_{i = 1}^{N}\vert e_1(i)\vert\,.
\end{equation}
\item Standard deviation performance index
\begin{equation}
\sigma_e = \sqrt{\frac{1}{N}\sum_{i =1}^N\left( \vert e_1(i)\vert - \mu_e \right)^2 }\,.
\end{equation}
\end{enumerate}
The maximal absolute value of the tracking error can be considered as an index of the tracking accuracy.  Meanwhile, the average tracking error and the standard deviation performance show the average tracking performance and the deviation of the tracking errors, respectively. Finally, the Integral Square Error (ISE) will be taken into account.

The experimentally obtained values for every index are presented in table \ref{tab:table_index}, taking the error in the last step response, that is, from $t = 10$ to $t = 14$.
\begin{table}[h!]
  \begin{center}
     \caption{Values of perfomance indices}
    \label{tab:table_index}
    \begin{tabular}{l|c} 
      \hline\hline
      \textbf{Index:} & \textbf{Value:} \\
      \hline\hline
      $M_e$ & 0.0064 \\ \hline
      $\mu_e$ & 0.0011 \\ \hline
      $\sigma_e$ & 3.4598 $\times 10^{-6}$ \\ \hline
       ISE & $333.7830$  \\ \hline
       Percentage of steady state error & $0.55\%$  \\ \hline
    \end{tabular}
  \end{center}
\end{table}
The percentage of the steady state error is obtained taking the average tracking error as a percentage of the total cylinder drive, that is $0.20$ m.
\section{Conclusion}
\label{sec:conclusions}

A new practical control approach, based on the sliding mode STA,
is introduced and applied to a hydraulic cylinder actuator. The
proposed method is proved to track a smooth trajectory in presence
of both matched and unmatched perturbations, producing an
acceptably small and ultimately bounded tracking error. Important
to notice is that via the designed integral sliding surface, the
velocity state is fully excluded and needs to be neither measured
nor estimated for implementing the proposed controller. The design
methodology for such surface is developed in terms of an LMI. The
feasibility of the proposed approach is shown via experimental
results achieved on a standard industrial hydraulic actuator in
laboratory setting, without additional measurements and extensive
modeling and identification.

\begin{ack}
Leonid Fridman and Manuel A. Estrada are grateful for the
financial support of CONACyT (Consejo Nacional de Ciencia y
Tecnolog\'ia): CVU 833748; PAPIIT-UNAM
(Programa de Apoyo a Proyectos de Investigaci\'on e Innovaci\'on
Tecnol\'ogica): Project IN106622.
\end{ack}

\bibliographystyle{plain}
\bibliography{ifachydrabib}             

\appendix
\section{Proof of Theorem 1}    
Assume the following Lyapunov function candidate $V(e) = e^T M e$.
Its derivative along the trajectories of $\dot{e} =( A_n - BR)e +
\psi(t,e)H e$ has the form
\begin{equation}
\begin{aligned}
\dot{V}(e) &= e^T(t) \left[  M A_n + A_n^TM + \psi(t,e)\left( H^TM + MH\right) \right]e(t) \\& -  e(t)^T\left(K^TB^TM +MBK\right)e(t)  \\
&\leq e^T(t) \left[ MA_n+A_n^TM -K^TB^TM - MBK \right. \\ & \left.+\Psi \left( H^TM + MH\right) \right]e(t)
\end{aligned}
\end{equation}
Then, as shown in \cite{Corless1994}, the system is quadratically
stable if there exists such $R\in \mathbb{R}^{3}$ and $M=M^T>0$
that \eqref{eq:LyapIneq} is satisfied. The necessity comes from
the quadratic stabilizability. If the system is not quadratically
stabilizable, then there is no solution of \eqref{eq:LyapIneq}.
This concludes the proof.

\section{Proof of Lemma 3}
If $\beta(t) \neq 0$, the closed loop of mechanical sub-system
with the virtual control has the form $\dot{e}(t) = (A_n -BK)e +
\psi(t,x)He + \beta(t)D$, where $D^T =\begin{bmatrix} 0 & 1 & 0
\end{bmatrix}$.  Taking the Lyapunov function candidate $V =
e^TMe$, its derivative is
\begin{equation*}
\begin{aligned}
\dot{V}(e) &= e^T(t) \left[  M A_n + A_n^TM + \psi(t,e)\left( H^TM + MH\right) \right]e(t) \\& -  e(t)^T\left(K^TB^TM +MBK\right)e(t) \\ & + \beta(t)\left( D^TMe + e^TMD  \right)  \\
&\leq e^T(t) \left[ MA_n+A_n^TM -K^TB^TM - MBK \right. \\ & \left.+\Psi \left( H^TM + MH\right) \right]e(t)  + 2\lambda_{max}(M)\beta_M\Vert e(t)\Vert\,.
\end{aligned}
\end{equation*}
If the LMI \eqref{eq:LyapIneq} is satisfied, then there exists
$\mu(M)>0$ such that
\begin{equation}
\dot{V}(e) \leq - \mu(M)\Vert e(t)\Vert^2 + 2\lambda_{max}(M)\beta_M\Vert e(t)\Vert
\end{equation}
Assuming a constant $0<\theta_V<1$, the following is satisfied
\begin{equation*}
\dot{V}(e) \leq -(1-\theta_V)\mu(M)\Vert e(t)\Vert^2\,,\, \Vert e(t)\Vert \geq \frac{2\lambda_{max}(M)\beta_M}{\theta_V\mu(M)}\,.
\end{equation*}
By Theorem 4.18 from \cite{Khalil2001}, the system's trajectories
are then ultimately uniformly bounded and the bound
\eqref{eq:ultimatebound} is satisfied.

\section{Proof of Lemma \ref{lem:reach} }
In order to show the reachability of the sliding surface \eqref{eq:surface}, consider the following control
\begin{equation}
\begin{aligned}
    u(t) =& - \frac{V_t}{4\tau E C_q}K_s \text{sign}(s) \\ &+ \frac{V_t}{4\tau E C_q}\left[\gamma_1e_1 - \left(\gamma_2 + \frac{4EC_{qp}}{V_t}\right)\eta\right]
    \end{aligned}
\end{equation}
for a positive constant $K_s$. Therefore, the reachability condition has the form:
\begin{equation}
    \begin{aligned}
s\dot{s} &= s\left[ -K_s\text{sign}(s) + \bar{\delta}_3(t) + \kappa e_2 \right]\\
&\leq -K_s\vert s\vert + \vert s\vert \vert \bar{\delta}_3(t)\vert + \vert s\vert \kappa\vert e_2\vert \\
& \leq -\vert s\vert \left( K_s + L_3 + \kappa\vert e_2\vert \right)\,,
    \end{aligned}
\end{equation}
since $\vert e_2\vert \leq C_{e_2}$,  there always exist $K_s > L_3 + \kappa C_{e_2} $, such that with $\bar{K}_s= K_s - L_2 -\kappa C_{e_2} $, and the following is satisfied:
\begin{equation}
    s\dot{s} \leq -\bar{K}_s\vert s\vert <0\,,
\end{equation}
proving that there always exist a control law (indepedent of the uncertainties and perturbations) that ensures the trajectories converges to the manifold $s(t) \equiv 0$, and therefore, they remain in such manifold.
\section{Proof of Lemma \ref{lemma:sta} }              
To prove the stability of $s = 0$ and $ z = 0$, consider the
following change of coordinates introduced in \cite{Moreno2012}:
$\zeta^T = \begin{bmatrix} \zeta_1 & \zeta_2 \end{bmatrix} = \begin{bmatrix} \lceil s\rfloor^{1/2} & z
\end{bmatrix}$. Computing its dynamics yields
\begin{equation}
\dot{\zeta} = \frac{\partial \zeta_1}{ \partial s} \begin{bmatrix}
-k_1\rho & 1 \\ -k_2\rho^2 & 0
\end{bmatrix}\zeta+ \begin{bmatrix}
0 \\ \delta_z(t,e)
\end{bmatrix} = \frac{\partial \zeta_1}{ \partial s} A_s\zeta + G\delta_z(t,e)\,,
\end{equation}
where $G^T = \begin{bmatrix} 0 & 1 \end{bmatrix}$. Note that this term is
always positive and bounded from below as $\left\vert \frac{\partial \zeta_1}{ \partial s} \right\vert
\geq 1$.  Consider the following scaling of the states
\begin{equation}
\xi = \begin{bmatrix}
\xi_1 \\ \xi_2
\end{bmatrix} = \Gamma^{-1}\zeta \quad \text{with}\quad \Gamma =
\begin{bmatrix} \rho & 0 \\ 0 & \rho^2
\end{bmatrix}\,,
\end{equation}
so that the dynamics has the form $\dot{\xi} = \frac{\partial \xi_1}{ \partial s}
\Gamma^{-1}M_s\xi + \Gamma^{-1}G\delta_z(t,e)$ . Then one has
\begin{equation}
\frac{d \xi}{dt} = \frac{\partial \xi_1}{ \partial s} \left( \rho A_k \xi + G\frac{\delta_z(t,e) }{ \frac{\partial \xi_1}{ \partial s}\rho} \right)\,.
\end{equation}
Consider the Lyapunov function candidate $V_s = \xi^TM_k\xi$,
whose derivatives along the trajectories of the $\dot{\xi}$ vector
field has the form
\begin{equation}
\begin{aligned}
\frac{d V_s(t)}{dt} &= \dot{\xi}^TM_k \xi + \xi^TM_k \dot{\xi}\\
=& \rho  \frac{\partial \xi_1}{ \partial s}  \xi^T\left( A_k^TM_k+ M_kA_k\right) + 2 \frac{\partial \xi_1}{ \partial s}
\xi^TM_kG \frac{\delta_z(t,e)}{ \frac{\partial \xi_1}{ \partial s} \rho}\,.
\end{aligned}
\end{equation}
Note that for any Hurwitz matrix $A_k$, there exists always a
solution $M_k$ of $A_k^TM_k+ M_kA_k = -\mathbb{I}$, so that
\begin{equation}
\frac{d V_s(t)}{dt} = - \frac{\partial \xi_1}{ \partial s} \left(  \rho \xi^T\xi -  2
\xi^TM_k G \frac{\delta_z(t,e)}{ \frac{\partial \xi_1}{ \partial s} \rho}\right)\,.
\end{equation}
For the assumed $\vert \delta_z\vert \leq L \vert \xi_1\vert$, one
can obtain the following bound
\begin{equation}
\frac{d V_s(t) }{dt} \leq  -  \frac{\partial \xi_1}{ \partial s} \left(  \rho  -  2 L
\lambda_{max}(M_k)  \right)\Vert \xi\Vert^2\,.
\end{equation}
It is sufficient that only the gain $\rho$ satisfies
\eqref{eq:rho}. Then, the stability of the surface's origin and
control law of the closed loop are ensured for a sufficiently
large value of $\rho$.

\end{document}